\documentclass[preprint,preprintnumbers,amsmath,amssymb]{revtex4}

\usepackage{graphicx}
\usepackage{txfonts}
\usepackage{type1cm}
\usepackage{amsmath, amssymb}
\usepackage{braket}
\usepackage{bm}
\usepackage{xcolor}
\usepackage{ulem}

\DeclareRobustCommand{\er}{\bgroup\markoverwith{\textcolor{red}{\rule[.5ex]{2pt}{0.4pt}}}\ULon} 

\begin{document}

\title{
 {\bf 
  Determination of the magnetic $\bm q$ vectors in the heavy fermion superconductor Ce$_3$PtIn$_{11}$ 
 }
}

\author{
 Naoki Shioda$^{1}$, 
 Kazuki, Kumeda$^{1}$, 
 Hideto Fukazawa$^{1}$, 
 Tetsuo Ohama$^{1}$, 
 Yoh Kohori$^{1}$, 
 Debarchan Das$^{2, 3}$, 
 Joanna B{\l}awat$^{2}$, 
 Dariusz Kaczorowski$^{2}$, 
 Koudai Sugimoto$^{4}$
}

\address{
 $^1$Department of Physics, Chiba University, Chiba 263-8522, Japan\\
 $^2$Institute of Low Temperature and Structure Research, Polish Academy of Science, P.O. Box 1410, PL-50-950 Wroc{\l}aw, Poland\\
 $^3$Laboratory of Muon Spin Spectroscopy, Paul Scherrer Institute, CH-5232 Villigen PSI, Switzerland \\
 $^4$Department of Physics, Keio University, Yokohama 223-8522, Japan
}

\begin{abstract}
 An analysis with transferred hyperfine field has been performed on the spectra of the $^{115}$In nuclear quadrupole resonance experiments of heavy fermion superconductor Ce$_{3}$PtIn$_{11}$ 
 exhibiting co-occurrence of two successive antiferromagnetic orderings ($T_{\rm N1}$ = 2.2 K and $T_{\rm N2}$ = 2.0 K) followed by superconducting transition ($T_{\rm c}$ = 0.32 K). 
 The spectral changes at magnetic transition temperatures $T_{\rm N1}$ and $T_{\rm N2}$ indicate that the Ce(2) site has the dominant magnetic contribution, with a small magnetic moment, but not negligible, at the Ce(1) site.
 Our analysis using transferred hyperfine field which overcomes the previous simple dipolar model evinces that for$T_{\rm N2} < T < T_{\rm N1 } $, the propagation vectors at the Ce(1) and Ce(2) sublattices are ${\bm q}_{1} = {\bm q}_{2} = \left( \frac{1}{2} ,~ \frac{1}{2} ,~ 0 ~ \text{ or } ~ \frac{1}{2} \right) $, 
 whereas for $T < T_{\rm N2 }$, the propagation vectors are ${\bm q}_{1} = \left( \frac{1}{2} ,~ \frac{1}{2} ,~ \frac{1}{6} ~ \text{or } ~ \frac{1}{3} \right) $ and ${\bm q}_{2} = \left( \frac{1}{2} ,~ \frac{1}{2} ,~ \frac{1}{2} ~ \text{ or } ~ 0 \right) $, respectively. 
\end{abstract}


\maketitle

\section{Introduction}


Quantum criticality and quantum phase transition in rare earth based strongly correlated systems is one of the intensively studied topics in contemporary condensed matter research due to many exotic ground state properties 
\cite{2005coleman, 2017belitz, 2010si, 2006park, 2019das_usa, 2017kaluarachchi}. 
In this regards, Ce$_{n} M_{m}$In$_{3n+2m}$($M$ = Co, Rh, Pd, Ir, Pt) systems turn out to be quite promising because of the interplay between heavy-fermion superconductivity and magnetism in close proximity with quantum critical point. 
\cite{2001petrovic, 2000hegger, 2001kawasaki, 2010Kaczorowski, 2012fukazawa, 2001kohori, 2007fukazawa, 2010yashima}. 
For instance, CeIn$_{3}$ is known to exhibit a high N{\'e}el temperature ($T_{\rm N} = 10 ~ {\rm K} $) which can be suppressed by applying external pressure leading to a superconducting dome 
\cite{1998mathur}. 
While Ce$M$In$_{5}$ systems have relatively high superconducting transition temperatures $T_{\rm c}$, Ce$_{2}M$In$_{8}$ system has relatively lower $T_{\rm c}$ than Ce$M$In$_{5}$ system, suggesting relatively weak antiferromagnetic fluctuations in Ce$_{2} M $In$_{8}$ 
\cite{2001monthoux}. 
In particular, these materials have generated interest because of their varied magnetic structures.
The magnetic structures of CeIn$_{3}$, CeRhIn$_{5}$, and Ce$_{2}$RhIn$_{8}$ have been investigated by neutron scattering experiments
\cite{1980benoit, 2001bao}. 
However, unless the magnetic structure is simple, its direct determination by neutron scattering is difficult because of the very high neutron absorption cross-section of the $^{115}$In nucleus, making such experiments time-consuming.
Therefore, indirect determination or deduction of the magnetic structures of these materials using nuclear magnetic resonance (NMR) or nuclear quadrupole resonance (NQR) is quite popular and widely accepted 
\cite{2011sakai, 2017raba, 2017gauthier, 2000curro, 2006curro}.


\begin{figure}[t]
 \includegraphics[width = 20 em ]{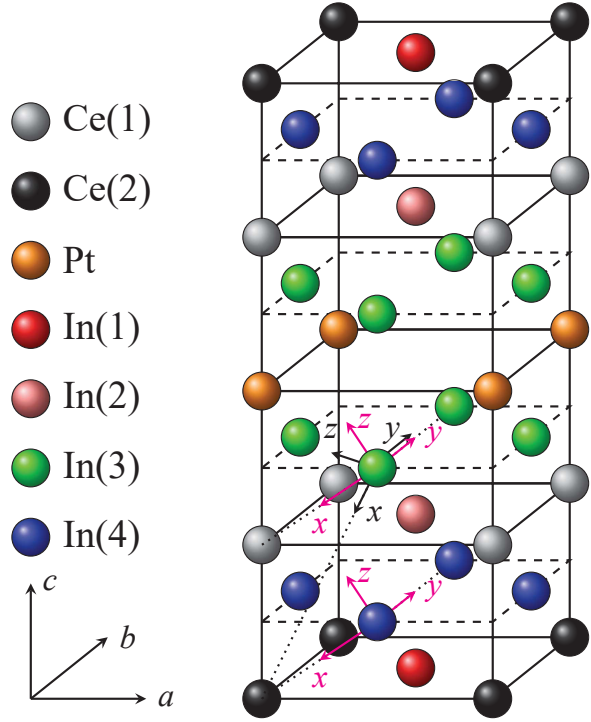}
 \caption{
  Crystal structure of Ce$_3$PtIn$_{11}$. 
  Arrows represent bond coordinates as specified in text. 
 }
 \label{kouzou}
\end{figure}


Ce$_3$PtIn$_{11}$ ($n = 3 ,~ m = 1 ,~ M = $ Pt) is a relatively new member of the Ce$_{n} M_{m}$In$_{3n + 2m}$ family
\cite{2014kratochvilova, 2018das}.
Fig. \ref{kouzou} shows the tetragonal crystal structure of Ce$_3$PtIn$_{11}$, which consists of periodically stacked two layers of CeIn$_3$ and one layer of CePtIn$_5$. 
In contrast with Ce$M$In$_{5}$, Ce$_{2}M$In$_{8}$, and CeIn$_{3}$, this material has two inequivalent Ce sites, Ce(1) and Ce(2), in the unit cell.
The Ce(1) site has a local 4$mm$ symmetry and an environment similar to that of the Ce site in Ce$_{2}$PtIn$_{8}$.
However, the Ce(2) site has a local 4/$mmm$ symmetry that is similar to that of the Ce site in CeIn$_3$.
Interestingly, Ce$_3$PtIn$_{11}$ undergoes successive magnetic transitions at ambient pressure ($T_{\rm N1} \simeq 2.2$ K, $T_{\rm N2} \simeq 2.0$ K) followed by superconducting transition at a low temperature ($T_{\rm c } = 0.32 $ K)
\cite{2014kratochvilova, 2015prokleska, 2018das}.
Such successive magnetic transitions have also been observed for Ce$_3$PdIn$_{11}$ ($T_{\rm N1 } \simeq 1.68 ~ {\rm K} ,~ T_{\rm N2 } \simeq 1.55 ~ {\rm K}$), which has the same crystal structure as Ce$_{3}$PtIn$_{11}$ 
\cite{2015kratochvilova, 2019das},
suggesting this is a common magnetic property of Ce$_3M$In$_{11}$ systems.
Temperature dependence of the magnetic entropy shows that the Ce(2) site mainly contribute to this antiferromagnetic phase, whereas the magnetic moment at the Ce(1) site is screened almost perfectly by the Kondo effect
\cite{2015prokleska}.
Therefore, it is important to consider the duality of the Ce sites in determination of the magnetic structure. 
Co-occurrence of antiferromagnetism and superconductivity has also been observed for CeRhIn$_{5}$ and Ce$_{2}$RhIn$_{8}$ ($ P \sim $1.6 GPa for CeRhIn$_{5}$, $ P \sim $1.3 GPa for Ce$_{2}$RhIn$_{8}$)
\cite{2003kawasaki, 2003nicklas}.
However, notably, for Ce$_{3}$PtIn$_{11}$, both the successive magnetic transitions and the superconducting transition occur at ambient pressure.
Hence, the duality of Ce sites may play an essential role in governing such a complex ground state exhibiting co-occurrence of AFM ordering and superconductivity. 



To determine the magnetic structure of this material, we prelously performed $^{115}$In-NQR experiments and conducted an analysis based on the magnetic dipolar model
\cite{2020fukazawa}. 
The results showed that it has a commensurate magnetic structure when $T_{\rm N1} < T < T_{\rm N2}$. 
In our previous work, we found that the Ce(1) site may possess a small magnetic moment, whereas the Ce(2) site has a significant magnetic moment. 
In addition, a qualitative magnetic structure was proposed for $T < T_{\rm N2}$. 
The structure is commensurate on the Ce(2) sublattice and incommensurate on Ce(1). 
However, we could not quantitatively reproduce the representative frequency evolution and the intensity ratio of the peaks using the magnetic dipolar model. 
To address these issues and to determine the magnetic structure of this compound with better resolution, we have carried out new sets of NQR measurements and performed spectral analysis beyond the dipolar model to reproduced the $^{115}$In-NQR spectra of Ce$_{3}$PtIn$_{11}$ with improved precision. 
In this study, we have obtained the magnetic structure of Ce$_{3}$PtIn$_{11}$ with improved precision by conducting analysis considering the transferred hyperfine field.


\section{Experimental Setup and Results}


A polycrystalline sample of Ce$_{3}$PtIn$_{11}$ was obtained using the two-stage arc melting method as described in detail in Ref 
\cite{2021das}.
The sample was grounded to powder to enhance the NQR intensity and reduce the heat generation.
$^{115}$In-NQR studies were conducted in the frequency range of 7--70 MHz using a phase-coherent pulsed NQR spectrometer.
The temperature was controlled using a $^{4}$He cryostat in the temperature range of 1.4 $< ~ T ~ < $ 4.2 K.



The $^{115}$In-NQR $\left( I = \frac{9}{2} \right) $ spectrum in the paramagnetic state is expected to mainly consist of four lines contributed by each of the four inequivalent In sites.
These lines are observed at $f = \nu_{Q},~2\nu_{Q},~3\nu_{Q},~4\nu_{Q}$ from the sites with $\eta = 0$, where $\eta$ is the asymmetry parameter of electric field gradient; however, they are complex owing to the low symmetry of the In(3) and In(4) sites (local $2mm$ symmetry for both sites).
Therefore, we denote the lines from each site as $\nu_{1},~\nu_{2},~\nu_{3},~\nu_{4}$, in ascending order of their frequencies.
But, in our previous study, only 9 of the 16 expected lines were observed: $\nu_{4}$ from In(1); $\nu_{2},~\nu_{3},~\nu_{4}$ from In(2); $\nu_{1},~\nu_{2},~\nu_{3},~\nu_{4}$ from In(3); and $\nu_{4}$ from In(4)
\cite{2020kambeprb, 2020fukazawa}.


\begin{figure}[t]

 \begin{tabular}{l}
  \includegraphics[height = 34 ex]{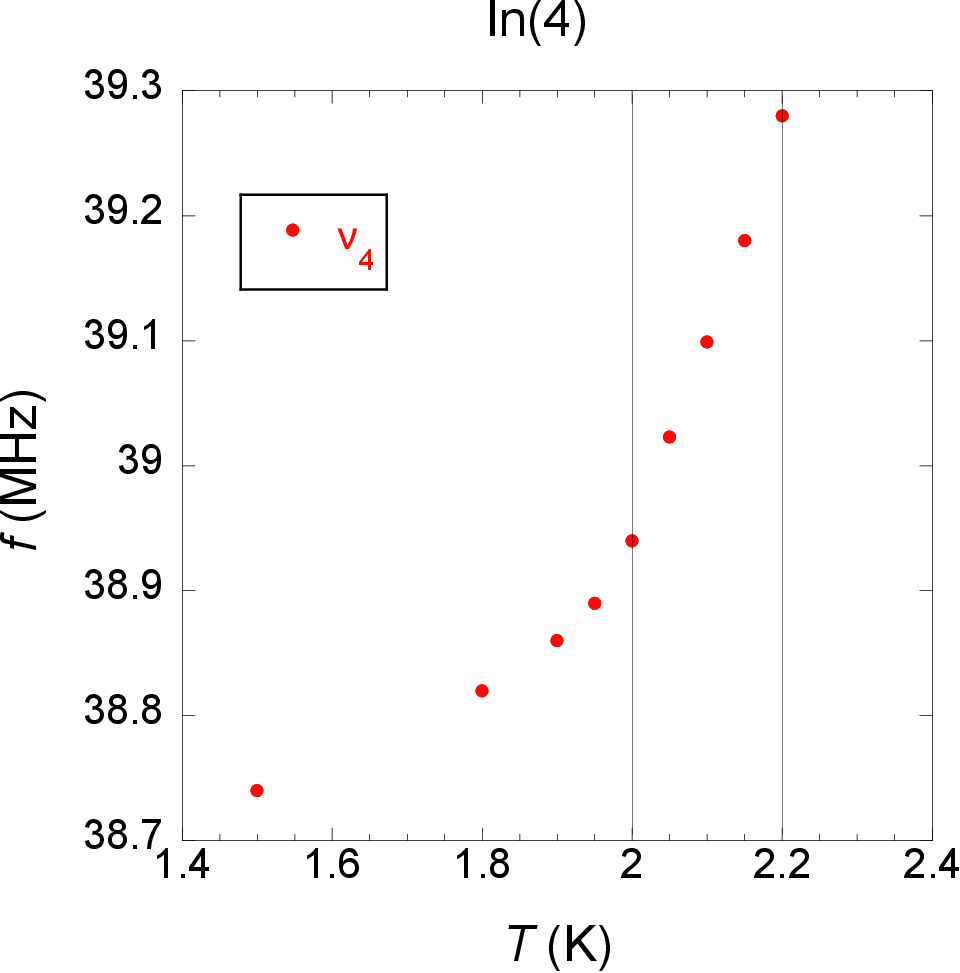} \\
  \includegraphics[height = 35 ex]{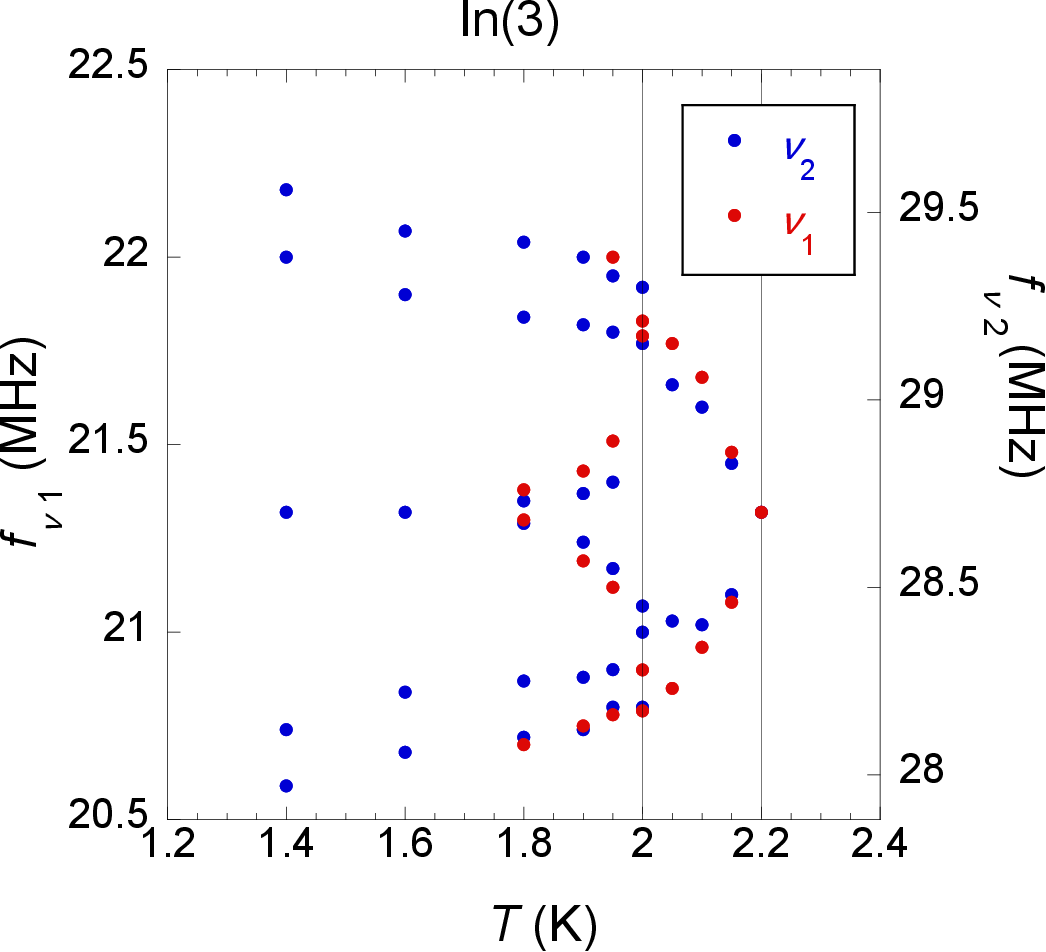} 
 \end{tabular}

 \caption{
  Temperature dependence of representative frequencies in spectra of Ce$_3$PtIn$_{11}$
  (a) In(4)-$\nu_4$ line measured in this study.
  (b) In(3)-$\nu_1$ and $\nu_2$ lines measured previously adapted from ref 
  \cite{2020fukazawa}. 
 }

 \label{tempdep}

\end{figure}

\begin{table}[b]

 \caption{
  Behavior of observed lines in NQR spectra of Ce$_3$PtIn$_{11}$ at temperatures $T_{\rm N 1} $ and $T_{\rm N 2} $.
 }

 \label{3S}

 \begin{tabular}{|c|c|c|c|} 
  \hline
  site & line & $T_{\rm N 1}$ & $T_{\rm N 2}$ \\
  \hline \hline
  In(1) & $\nu_4$ & No change & 
 Not observed \\
  \hline
   & $\nu_2$ & No change & No change \\
  \cline{2-4}
  In(2) & $\nu_3$ & No change & No change \\
  \cline{2-4}
   & $\nu_4$ & No change & No change \\
  \hline
   & $\nu_1$ & Split into 2 lines & Split into 4 lines \\
  \cline{2-4}
  In(3) & $\nu_2$ & Split into 2 lines & Split into 6 lines \\
  \cline{2-4}
   & $\nu_3$ & No change & No change \\
  \cline{2-4}
   & $\nu_4$ & No change & No change \\
  \hline
  In(4) & $\nu_4$ & Shift & Shift \\
  \hline
 \end{tabular}

\end{table}


However, in our previous study 
\cite{2020fukazawa}, 
the temperature dependence of the representative frequency of the In(4)-$\nu_{4}$ line 
was not smooth, particularly at approximately 2.0 K, probably owing to the inaccurate temperature control caused by the sample heating. 
Because of the weak signal of the In(4)-$\nu_{4}$ line, the excessive power of the NQR pulse injected into the sample could have reduced smoothness. 
In this study, we adjusted the temperature more accurately by lowing the pulse power and increasing the pulse interval, and consequently obtained a smooth result. 
The temperature dependencies of the representative frequencies of the In(4)-$\nu_{4}$ line obtained in the present study and the In(3) -$\nu_{1}$ and -$\nu_{2} $ lines obtained in the previous study
\cite{2020fukazawa}
are shown in Fig. \ref{tempdep}. 
The other observed lines, which were previously observed, were unchanged around $T_{\rm N1 } $ and $ T_{\rm N2 } $. 
The temperature dependencies of the representative frequencies in the NQR spectra around $T_{\rm N1}$ or $T_{\rm N2}$ were classified into three types: no change, shift, and split, as summarized in Table \ref{3S}.



Here, we discuss the possible magnetic structures from the spectral changes based on the dipolar model of Ref.
\cite{2020fukazawa}.
First, the lines from the In(2) (and In(1)) sites do not change; suggesting that the internal fields at these sites must be zero or very small.
As these sites are located at the center of the four Ce sites in the Ce(1) (or Ce(2)) plane of the unit cell, the magnetic structure, considering the internal fields at the In(2) and In(1) sites as zero is expressed as
\begin{align}
 &{\bm m}_{\kappa} (T) \parallel c \quad \quad \quad (\kappa = 1 ,~ 2 ) \label{mparac} \\
 &{\bm q}_{\kappa} = \left(
  \frac{1}{2} ,~ \frac{1}{2} ,~ q_{c ,~ \kappa}
 \right). \label{InPlaneAFM}
\end{align}
Here, $\kappa $ is the index of Ce sites, and ${\bm m}_{\kappa} $ and $ {\bm q }_{\kappa} $ are the magnetic moment and propagation vector of the Ce($\kappa $) site, respectively.
Magnetic structures such as the propagation vector in Eq. (\ref{InPlaneAFM}) are analogous to those of other Ce-based materials, e.g., 
CeIn$_{3}$ $\left( {\bm q} = \left( \frac{1}{2} ,~ \frac{1}{2} ,~ \frac{1}{2} \right) \right)$
\cite{1980benoit},
CeRhIn$_{5}$ $\left( {\bm q} = \left( \frac{1}{2} ,~ \frac{1}{2} ,~ 0.297 \right) \right)$
\cite{2000bao},
Ce$_{2}$RhIn$_{8}$ $\left( {\bm q} = \left( \frac{1}{2} ,~ \frac{1}{2} ,~ 0 \right) \right)$
\cite{2001bao},
and
CePt$_{2}$In$_{7}$ $\left( {\bm q} = \left( \frac{1}{2} ,~ \frac{1}{2} ,~ \frac{1}{2} \right) \right)$
\cite{2017raba, 2017gauthier}.

For the In(4)-$\nu_{4}$ line, a spectral shift was observed.
In this case, a single internal field perpendicular to the maximum principal axes of the electric field gradient (EFG), $V_{zz}$, is required.
By band calculations, it has been clarified that $V_{zz}$ at the In(4) site is perpendicular to the wall of the unit cell
\cite{2020fukazawa}.
For example, $V_{zz}$ at the In(3) or In(4) site located at $\left( \frac{1}{2} ,~ 0 ,~ c \right) $, as shown in Fig. \ref{kouzou}, is parallel to the $b$ axis. 
In the magnetic structure described above, the internal field at the In(4) nucleus is perpendicular to $V_{zz}$, which satisfies the requirement.
Because only a shift was observed at both $T_{\rm N1}$ and $T_{\rm N2}$ without any spectral broadening or splitting, the propagation vector at the Ce(2) sublattice, which is close to the In(4) site, must be commensurate; thus, 
\begin{align}
q_{c ,~ 2} = 0 ~{\rm or}~ \frac{1}{2}.
\label{IntraPlane}
\end{align}
Note that specific heat measurement indicated a large magnetic entropy release at $T_{\rm N2 } $.
It can be attributed to the magnetic moment at the Ce(2) site being much larger than that at the Ce(1) site. 
Therefore, the propagation vector at the Ce(2) sublattice changes at $T_{\rm N2}$, i.e., it corresponds to a commensurate-to-commensurate transition from $ \left( \frac{1}{2} ,~ \frac{1}{2} ,~ 0 \right) $ to $ \left( \frac{1}{2} ,~ \frac{1}{2} ,~ \frac{1}{2} \right) $ or vice versa.

Finally, regarding the In(3) site, spectral splitting was observed for the $\nu_{1}$ and $\nu_{2}$ lines.
Thus, an internal field perpendicular to $V_{zz}$ is expected, similar to the In(4) site case, which also satisfies the requirement described above.
Furthermore, these lines split into four and six lines, respectively, which suggested that the internal field was not single-valued, instead it had approximately three values. 
Therefore, the Ce(1) sublattice, which is close to the In(3) site, is expected to be an incommensurate structure with approximately three or six-fold period along the $k_{c} $ axis of the reciprocal lattice space.
However, this structure can not quantitatively explain these splittings when considering the magnetic dipolar model.
Another unresolved puzzle was that the extremely large magnetic moment was evaluated using this model ($| {\bm m}_{2} | \sim 7\mu_{\rm B}$) at the Ce(2) site.
This suggests that the essential contribution of the internal fields at the In nuclei is the coupling mediated by the conduction electrons, instead of direct magnetic dipolar interaction.


\section{Methods of Analysis}


The change in the NQR spectra of Ce$_3$PtIn$_{11}$ can be attributed to the internal fields due to the magnetic ordering.
The sum of the electric quadrupole Hamiltonian and the Zeeman Hamiltonian at the In($\lambda $) site ($\lambda = 1,~2,~3,~4$) is given by 
\begin{align}
 \mathcal{H} &= \mathcal{H}_{Q} +\mathcal{H}_{\rm Z} \nonumber \\
 &= \frac{e^2 qQ}{4I(2I-1)}
 \left\{
  (3 {I_z}^2 - I^2) + \frac{1}{2} \eta ( {I_+}^2 + {I_-}^2 )
 \right\}
 - \gamma \hbar {\bm I} \cdot {\bm H}^{\lambda}_{\rm int}, 
\end{align}
where $eq ~ ( = V_{zz} )$, $ eQ$, and $\eta ~ \left( = \frac{ | V_{xx} - V_{yy} | }{ V_{zz} } , ~ | V_{zz} | \geq | V_{yy} | \geq | V_{xx} | \right) $ are the EFG, nuclear quadrupole moment, and asymmetric parameter, respectively, and $\gamma$ is the gyromagnetic ratio.
The hyperfine field, ${\bm H}^{\lambda}_{\rm int}$, at the In($\lambda$) site is
\begin{align}
{\bm H}^{\lambda}_{\rm int} = \sum_{\kappa = 1 ,~ 2} \sum_{i \in{n.n.} } {\sf B}^{\lambda}_{\kappa ,~i} {\bm m}_{\kappa ,~i} ( T ),
\end{align}
where ${\sf B}^{\lambda}_{\kappa ,~i} $ is the transferred hyperfine tensor and $ m_{\kappa ,~i} ( T )$ is the magnetic moment at the nearest-neighboring Ce($\kappa$) sites.
There are four (for the In(1) and In(2) sites) or two (for the In(3) and In(4) sites) nearest-neighboring Ce sites.
The representative frequencies of the NQR spectra were analyzed by calculating the above Hamiltonian.



The interaction between the localized moment at the Ce site and the In nucleus is probably mediated by conduction electrons.
In the present analysis, the contributions, except of the nearest neighbors, are assumed to be renormalized in the hyperfine tensor. 
Subsequently, the calculations were performed under the following considerations:
the internal field of the nearest Ce(2) site for the In(1) and In(4) sites, and
those of the nearest Ce(2) and Ce(1) sites for the In(2) and In(3) sites.
These calculations of the hyperfine fields are based on those performed by Curro for the Ce$M$In$_{5}$ 
\cite{2006curro}. 



As shown in Fig. \ref{kouzou}, the bond coordinate system is defined with the $x$-axis in the bond direction of the In and Ce sites, $y$-axis in the $ab$-plane, and $z$-axis perpendicular to them.
Subsequently, we assume that the transferred hyperfine tensor, ${\sf B}^{\lambda}_{\kappa ,~i}$, in the bond coordinate system is diagonalized as
\begin{align}
 {\sf B} =\left(
 \begin{array}{ccc}
  B_{\parallel} & 0 & 0 \\
  0 & B_{\rm hori} & 0 \\
  0 & 0 & B_{\perp}
 \end{array}
 \right).
 \label{diagonal}
\end{align}
The hyperfine field is calculated by applying a coordinate transformation to Eq. (\ref{diagonal}). 

We focus on the In(1) and In(2) sites. 
As an example, consider the Ce(2) contribution to the In(2) site.
Let the Ce site located at $(a ,~ b ) = ( 1 ,~ 0 ) $ be $ i = 1 $ and the other nearest-neighboring Ce sites be numbered counterclockwise.
Consequently the transferred hyperfine tensor is
\begin{align}
 {\sf B}_{2 ,~ 1}^2 &= \left(
 \begin{array}{ccc}
  B_{2 ,~ \alpha}^2 & B_{2 ,~ \beta}^2 & B_{2 ,~ \gamma}^2 \\
  B_{2 ,~ \beta}^2 & B_{2 ,~ \alpha}^2 & B_{2 ,~ \gamma}^2 \\
  B_{2 ,~ \gamma}^2 & B_{2 ,~ \gamma}^2 & B_{2 ,~ \delta}^2
 \end{array}
 \right), \\
 {\sf B}_{2 ,~ 2}^2 &= \left(
 \begin{array}{ccc}
  B_{2 ,~ \alpha}^2 & -B_{2 ,~ \beta}^2 & B_{2 ,~ \gamma}^2 \\
  -B_{2 ,~ \beta}^2 & B_{2 ,~ \alpha}^2 & -B_{2 ,~ \gamma}^2 \\
  B_{2 ,~ \gamma}^2 & -B_{2 ,~ \gamma}^2 & B_{2 ,~ \delta}^2
 \end{array}
 \right), \\
 {\sf B}_{2 ,~ 3}^2 &= \left(
 \begin{array}{ccc}
  B_{2 ,~ \alpha}^2 & B_{2 ,~ \beta}^2 & -B_{2 ,~ \gamma}^2 \\
  B_{2 ,~ \beta}^2 & B_{2 ,~ \alpha}^2 & -B_{2 ,~ \gamma}^2 \\
  -B_{2 ,~ \gamma}^2 & -B_{2 ,~ \gamma}^2 & B_{2 ,~ \delta}^2
 \end{array}
 \right), \\
 {\sf B}_{2 ,~ 4}^2 &= \left(
 \begin{array}{ccc}
  B_{2 ,~ \alpha}^2 & -B_{2 ,~ \beta}^2 & -B_{2 ,~ \gamma}^2 \\
  -B_{2 ,~ \beta}^2 & B_{2 ,~ \alpha}^2 & B_{2 ,~ \gamma}^2 \\
  -B_{2 ,~ \gamma}^2 & B_{2 ,~ \gamma}^2 & B_{2 ,~ \delta}^2
 \end{array}
 \right), 
\end{align}
where
\begin{align}
 B_{2 ,~ \delta}^2 &= B_{2 ,~ \parallel}^2 + B_{2 ,~ \perp}^2 / 2,\notag \\
 B_{2 ,~ \alpha}^2 &= B_{2 ,~ \delta}^2 + B_{2 ,~ \rm hori}^2 / 2 ,\notag \\
 B_{2 ,~ \beta}^2 &= B_{2 ,~ \delta}^2 - B_{2 ,~ \rm hori}^2 / 2 ,\notag \\
 B_{2 ,~ \gamma}^2 &= B_{2 ,~ \parallel}^2 - B_{2 ,~ \perp}^2 / 2 \sqrt{2} . \notag
\end{align} 
In this case, the internal field is zero only if ${\bm m}_{2 ,~ 1} = - {\bm m}_{2 ,~ 2} = {\bm m}_{2 ,~ 3} = - {\bm m}_{2 ,~ 4} = {\bm m}$ and ${\bm m} \propto \hat{\bm c}$. 
This means that the magnetic moments at the Ce(2) sites neighboring in $ab$-plane are opposite. 
The same result also holds for Ce(1) to In(2) and Ce(2) to In(1). 
Therefore, it is confirmed that ${\bm q}_{\kappa} = \left( \frac{1}{2} ,~ \frac{1}{2} ,~ q_{c ,~ \kappa} \right) $ and $ {\bm m }_{\kappa}$ is parallel to $\hat{\bm c}$.

Hereafter, we consider the hyperfine magnetic fields at the In(3) and In(4) sites where spectral changes occurred.
The internal fields for the magnetic structure given by Eqs. (\ref{mparac})--(\ref{InPlaneAFM}) are calculated as 
\begin{align}
 {\bm H}^{3}_{\rm int} &= 
 \left(
  2 A_{1}^3 m_{1} (T) \cos ( 2 \pi q_{c ,~ 1} z + \delta ) + \frac{6}{5} A_{2}^3 m_{2} (T) \cos ( 2 \pi q_{c ,~ 2} z )
 \right) \hat{\bm c}, \\
 {\bm H}^{4}_{\rm int} &= 2 A_{2 }^4 m_{2} (T) \cos ( 2 \pi q_{c ,~ 2} z ) \hat{\bm c} ,\\
 A_{\kappa }^{\lambda} &\equiv \frac{ B_{\kappa ,~ \parallel}^{\lambda} - B_{\kappa ,~ \perp}^{\lambda} }{ 2 }.
\end{align}
In this model, $q_{c ,~ 2} = 0 ~ {\rm or } ~ \frac{1}{2} $ is again necessary to explain the shift of the In(4)-$\nu_{4}$ line.
Note that we considered the phase difference, $\delta$, between the magnetic structures of the Ce(1) and Ce(2) sublattices, which we did not consider in the magnetic dipolar model. 
This parameter $\delta$, was phenomenologically introduced due to existence of two Ce sites. 
We also assumed that the hyperfine tensor is independent of temperature and that the temperature dependencies of the representative frequencies are proportional to the evolution of the magnetic moments, $m_{\kappa} (T) $.
The temperature dependence of $m_{\kappa} (T) $ is given by the following localized magnetic model for the pseudospin, $S = \frac{1}{2}$:
\begin{align}
 m_{\kappa} ( T ) =  \mu_{ {\rm eff } ,~ \kappa} B_{S = \frac{1}{2} } 
 \left(
  \frac{ T_{\rm N} }{T}
  \frac{ 2S }{ S+1 }
  \frac{ m_{\kappa} ( T ) }{ \mu_{ {\rm eff } ,~ \kappa} }
 \right), 
 \label{Brillouin}
\end{align}
where $\mu_{ {\rm eff} ,~ 2} $ is $2.54 \mu_{\rm B } $ for $J = \frac{7}{2} $.
$\mu_{ {\rm eff} ,~1} = \left( m_{1} (T = 0) \right) $ is taken as a parameter because the magnetic moment is very small or zero. 
Therefore, we performed a fitting with the following five parameters: 
\begin{align}
q_{ c ,~ 1 } ,~ 
\delta ,~  
A_{ 2}^{ 4 } = A_{ 1 }^{ 3 } ,~ 
A_{ 2}^{ 3 } ,~ 
\text{and} ~
m_{ 1 } (T=0). \notag
\end{align}
Here, we set $A_{ 2}^{ 4 } = A_{ 1 }^{ 3 }$ to eliminate one extra degree of freedom because the distance between the Ce(2) and In(4) sites is similar to that between the Ce(1) and In(3) sites 
(
$d_{\rm Ce(2)-In(4) } / d_{\rm Ce(1)-In(3) } \simeq 1.0300$
)
\cite{2021das}.


\begin{figure*}[t]
 \begin{tabular}{cc}
  \includegraphics[width = 17 em ]{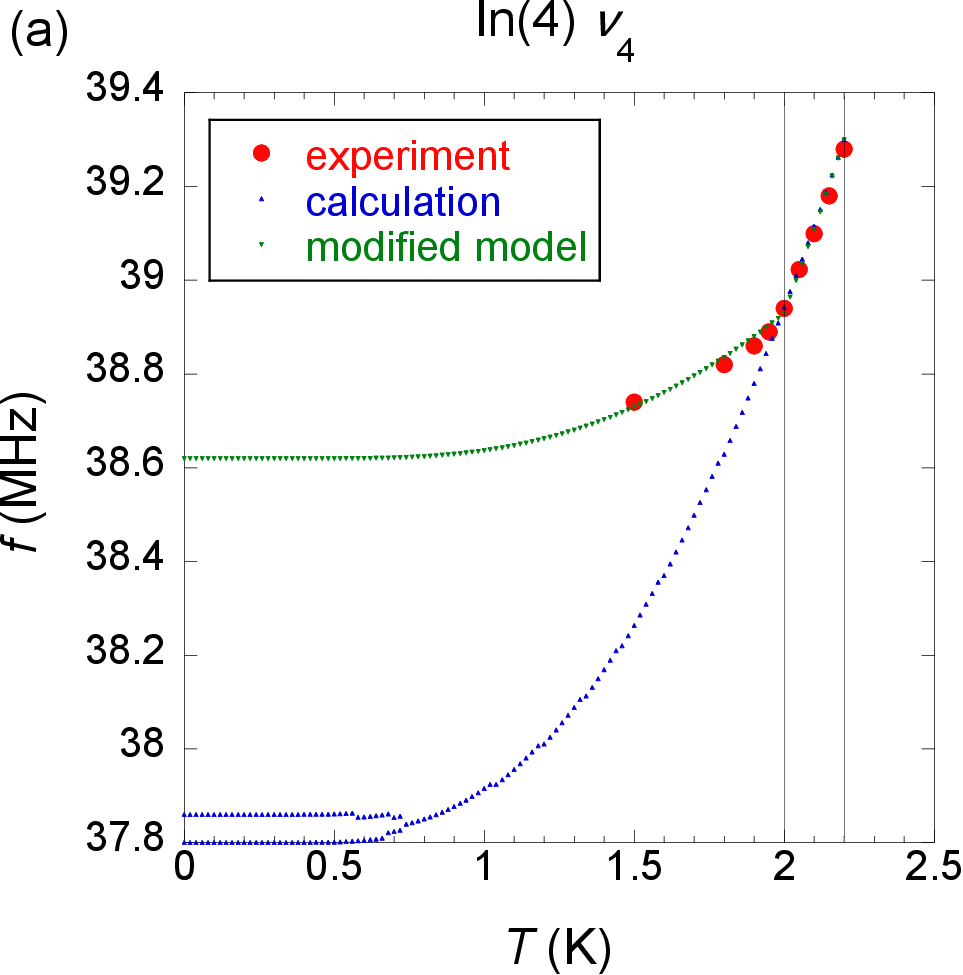} &
  \includegraphics[width = 17 em ]{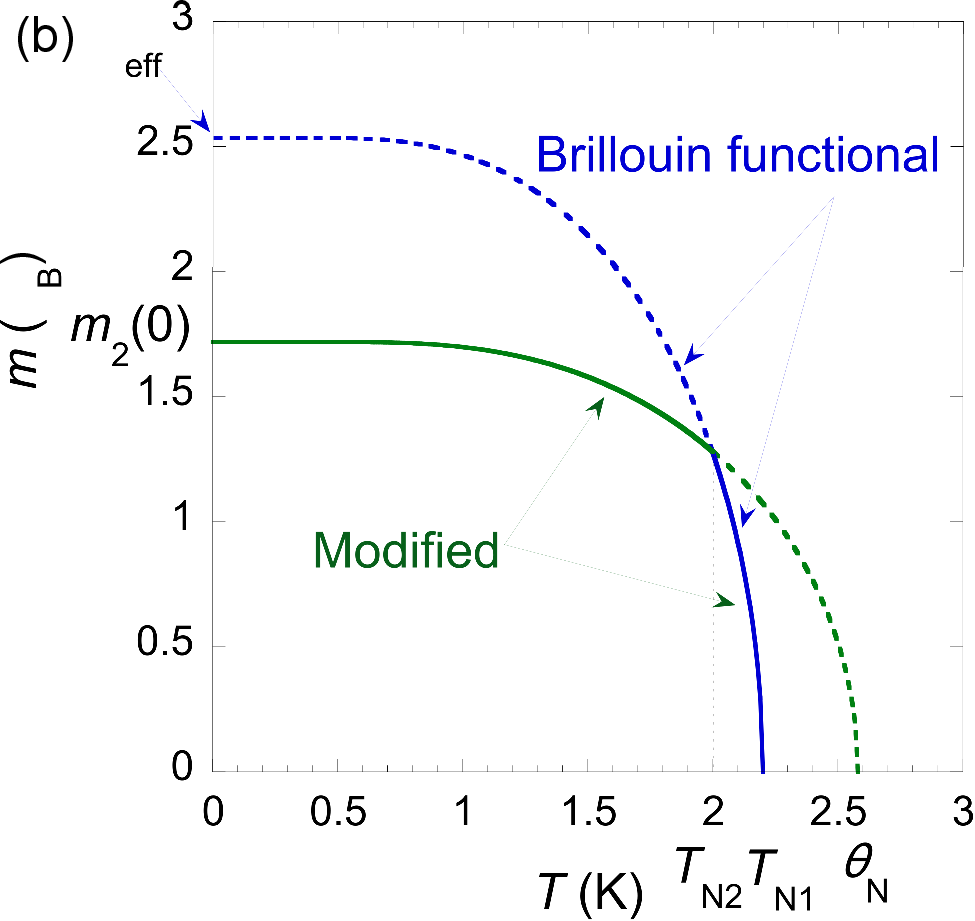} 
 \end{tabular}
 \caption{
  (a) Fitting results of In(4) $\nu_{4}$ line. 
  Blue curve is of simple Brillouin functional model, and green one is of modified model (shown in (b)). 
  (b) Temperature dependence model of ${\bm m}_{2}$. 
  Solid blue and dashed blue lines gives simple Brillouin functional model, whereas solid blue and solid green lines gives modified model. 
 }
 \label{In(4)}
\end{figure*}

\section{Results and Discussion of Analysis with Transferred Hyperfine Fields Model}

\begin{figure*}[t]
 \begin{tabular}{cc}
  \includegraphics[width = 17 em]{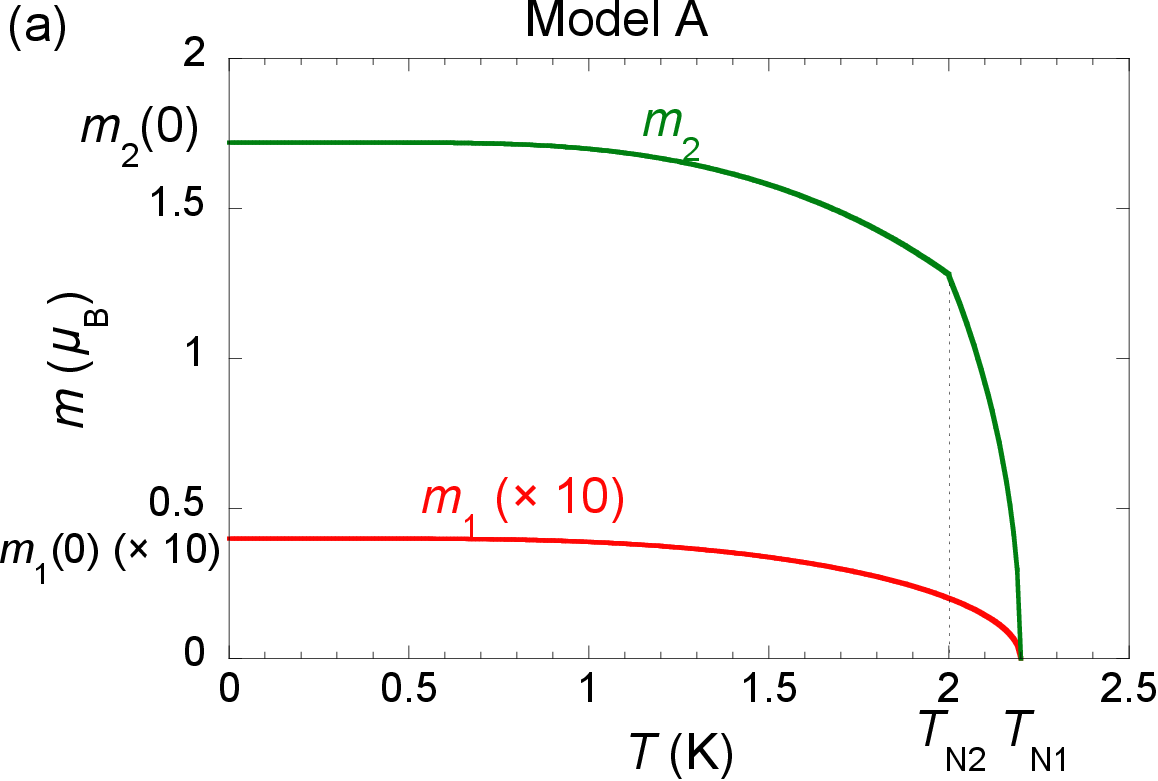} &
  \includegraphics[width = 17 em]{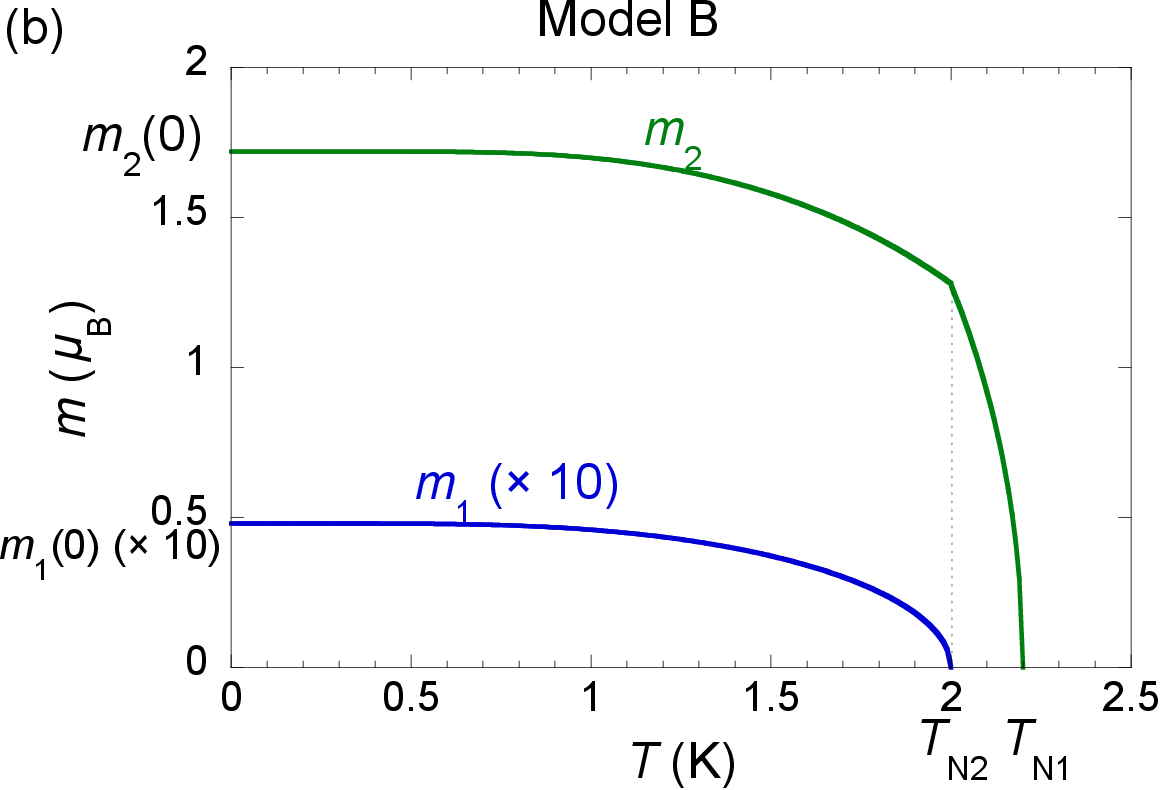} \\
  \includegraphics[width = 17 em]{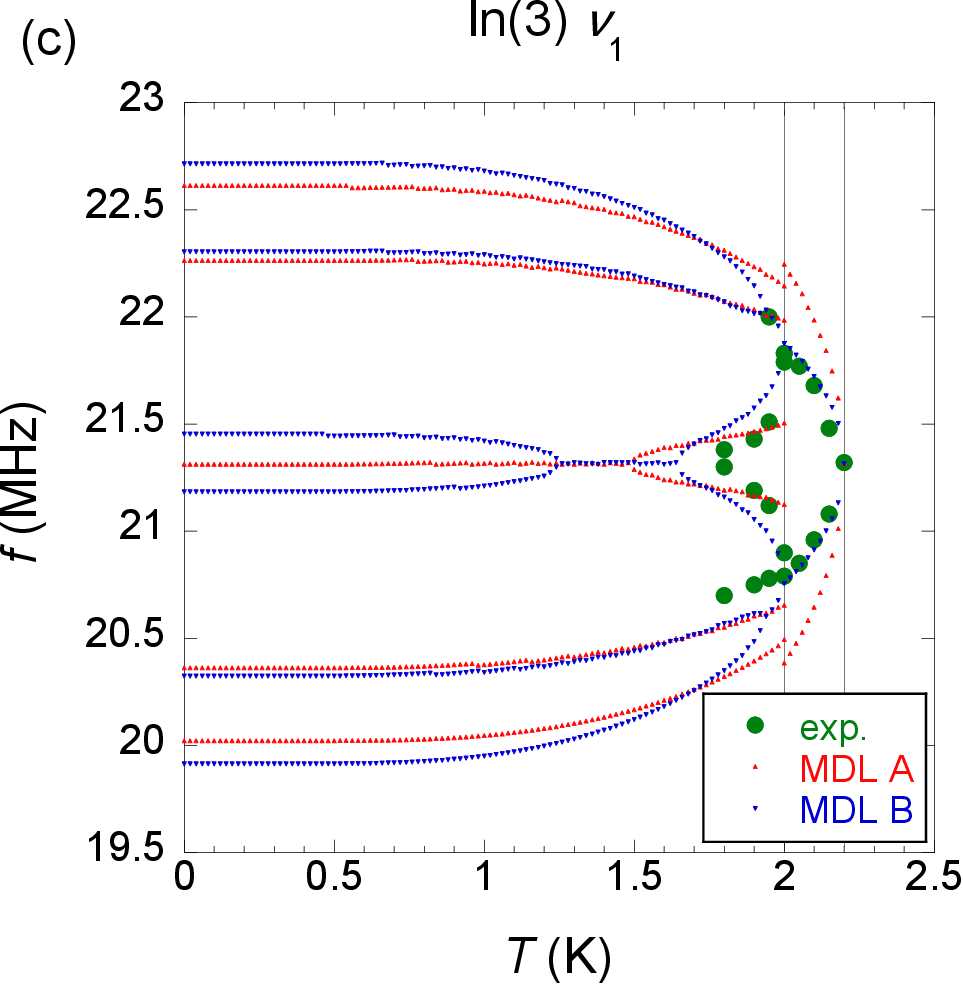} &
  \includegraphics[width = 17 em]{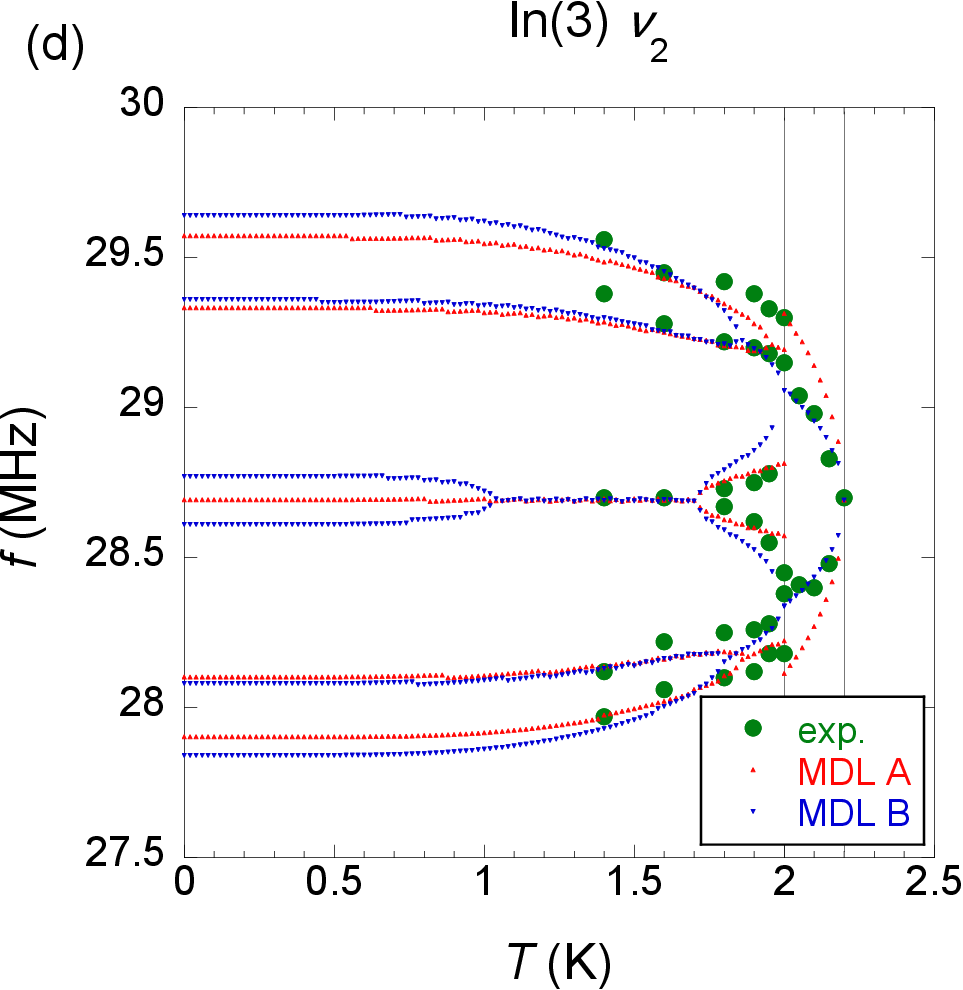} 
 \end{tabular}
 \caption{
  Two models, model A (MDL A) and model B (MDL B), of temperature dependence of magnetic moment at Ce(1) [(a), (b)] and their results for In(3) $\nu_{1} $ and $\nu_{2}$ lines [(c), (d)].
  Magnetic structures are 
  ${\bm q}_{2} ( = {\bm q}_{1} ) = \left( \frac{1}{2} ,~ \frac{1}{2} ,~ 0 \right) ,~ \delta = 0 $ for $T_{\rm N2} < T < T_{\rm N1}$ and
  ${\bm q}_{2} = \left( \frac{1}{2} ,~ \frac{1}{2} ,~ \frac{1}{2} \right) $ and ${\bm q}_{1} = \left( \frac{1}{2} ,~ \frac{1}{2} ,~ \frac{1}{6} \right) $ for $ T < T_{\rm N2}$. 
 }
 \label{In(3)}
\end{figure*}


First, we discuss the $\nu_{4}$ line of the In(4) site.
The red plot in Fig. \ref{In(4)} (a) shows the fitting result obtained using Eq. (\ref{Brillouin}).
This fit agrees well with the experimental results in the intermediate-temperature region $(T_{\rm N2} < T < T_{\rm N1})$, whereas it significantly deviates from those in the low-temperature region $(T < T_{\rm N2})$.
This suggests that the magnetic moment at the Ce(2) site below $T_{\rm N2}$ is smaller than the expected value for the localized magnetic model (2.54$\mu_{\rm B}$).
Therefore, we propose a phenomenologically modified model, as shown in Fig. \ref{In(4)} (b), where $\mu_{ {\rm eff} ,~ 2 }$ in the Brillouin function (Eq. (\ref{Brillouin})) is replaced by a new parameter $m_{2} (0) $ and a virtual N\'{e}el temperature $\theta_{\rm N} ( \simeq 2.6 ~ \text{ K } )$.
In this modified model, $m_{2} (T=0)$ is $1.72 \mu_{\rm B}$ and is less than the initial value, $\mu_{\rm eff} = 2.54, \mu_{\rm B}$ at $T = 0$.
This is in contrast with the magnetic dipolar model, which is unable to explain the extremely large magnetic moment at the Ce(2) site ($\sim 7 \mu_{\rm B} $). 
The results of the modified model are shown in blue color in Fig. \ref{In(4)} (a).
The plot reproduces the cusp-like evolution at $T_{\rm N2}$.
We use this modified temperature dependence of the magnetic moment at the Ce(2) site in the in our analysis.



Next, we discuss the $\nu_{1}$ and $\nu_{2}$ lines of the In(3) site.
In the intermediate-temperature region $(T_{\rm N2} < T < T_{\rm N1})$, because these lines split into two lines, the In(3) nucleus experiences an single-valued internal field.
Therefore, the magnetic structure of the Ce(1) sublattice must be commensurate with $q_{1} = 0 ~{\rm or }~ \frac{1}{2} $ or the magnetic moment must be $m_{1} (T) = 0$.
In the low-temperature region ($T < T_{\rm N2}$), the internal fields have several values, because the $\nu_{1}$ line splits into four and the $\nu_{2}$ line into six.
Therefore, $q_{1}$ must be incommensurate with a several-fold period along the $k_{c}$ axis in the reciprocal space.
Furthermore, some split lines merge into one line in both the $\nu_{1}$ and $\nu_{2} $ lines, which suggests the existence of In(3) nuclei with zero internal fields.
Hence, there must be opposite magnetic moments at the neighboring Ce(1) and Ce(2) sites.
As an example of such a magnetic structure, we propose the following parameters: $q_{c ,~ 1} = \frac{1}{6} ,~ q_{c ,~ 2} = \frac{1}{2} $ and $\delta = 165^{\circ}$.
Specific heat measurement suggests that the magnetic structure of the Ce(2) site is responsible for the magnetic transition at $T_{\rm N2} $
\cite{2015prokleska}
; hence, ${\bm q}_{2} = \left( \frac{1}{2} ,~ \frac{1}{2} ,~ 0 \right) ( = {\bm q}_{1})$ is required in the intermediate-temperature region.
An equivalent fitting is also obtained for another magnetic structure in the low-temperature region:${\bm q}_{1} = \left( \frac{1}{2} ,~ \frac{1}{2} ,~ \frac{1}{3} \right) ,~ {\bm q}_{2} = \left( \frac{1}{2} ,~ \frac{1}{2} ,~ 0 \right) ,~ \delta = 165^{\circ}$; in this case, ${\bm q}_{2} = \left( \frac{1}{2} ,~ \frac{1}{2} ,~ \frac{1}{2} \right) ( = {\bm q}_{1})$ is required in the intermediate-temperature region.
In addition, we propose two models for the temperature dependence of $m_{1} (T) $.
Both are Brillouin-functional models: in model A, the magnetic moment evolves from $T_{\rm N1 }$ (Fig. \ref{In(3)} (a)), and in model B, it grows from $T_{\rm N2 }$ (Fig. \ref{In(3)} (b)).
Fig. \ref{In(3)} (c) and (d) show the fitting results of the $\nu_{1}$ and $\nu_{2}$ lines of the In(3) site with each model, respectively.
Both reproduce the splittings and merges of the lines well, particularly in the low-temperature region.
However, in model A, the splitting is extremely large in the intermediate-temperature region, which suggests that a Brillouin-functional evolution does not occur at $T_{\rm N1 }$.
In model B, there is a difference between the fitting and experimental results immediately below $T_{\rm N2} $, which suggests a more rapid evolution of the magnetic moment or a discontinuous temperature dependence of $m_{1} (T) $.
The ratio of the magnetic moments at the Ce(1) and Ce(2) sites at $T = 0 $ is $\frac{m_{1}}{m_{2}} \simeq \frac{1}{40} $ in both models, which indicates that the magnetic moment at the Ce(1) site is very small even in the present model, similar to the previous magnetic dipolar model
\cite{2020fukazawa}.



It is worth discussing the order of the phase transition.
In this analysis, we assume second-order transitions at both $T_{\rm N1} $ and $T_{\rm N2}$.
However, a cusp-like temperature dependence is observed for the In(4)-$\nu_{4}$ line at $T_{\rm N2 }$, and the spectra of the In(3) site shows a discontinuous change at the same temperature.
The specific heat measurements reveal that at high magnetic fields ($\sim$ 3.5 T), $T_{\rm N1}$ and $T_{\rm N2}$ merge and become a single magnetic transition, which resembles to  the expected feature for a first-order transition
\cite{2018das}. 
On the other hand, a detailed investigation of this feature in Ce$_{3}$PdIn$_{11}$, which has the same crystal structure as Ce$_{3}$PtIn$_{11}$, do not exhibit clear latent heat and are suggested as second-order transitions
\cite{2019das}. 
Moreover, the metamagnetic transition 
\cite{2018das, 2019das} 
observed at higher magnetic fields ($\sim$ 5 GPa) is clearly first-order transition in nature. 
From the viewpoint of NQR, we cannot exclude the possibility of a first-order magnetic transition at $T = T_{\rm N2}$.


\section{Summary}


We analyzed the temperature dependence of the NQR spectra of Ce$_3$PtIn$_{11}$ by considering the transferred hyperfine field.
Because of temperature-independent lines, both the Ce(1) and Ce(2) sites must have ${\bm m}_{\kappa} \parallel c ,~ {\bm q}_{\kappa} = \left( \frac{1}{2} ,~ \frac{1}{2} ,~ q_{c ,~ \kappa} \right) $, and $(\kappa = 1 ,~ 2 ) $.
Thus, because the $\nu_{4} $ line of the In(4) site only shifts without any splitting or broadening below $T_{\rm N1} $, $q_{c ,~ 2} = 0 ~ {\rm or } ~ \frac{1}{2} $ is also required in this model.
Furthermore, we fitted for the $\nu_{1}$ and $\nu_{2} $ lines of the In(3) site with the above propagation vectors and several calculation parameters.
Thus, the temperature dependence of the representative frequencies below $T_{\rm N2} $ was reproduced by a magnetic structure with a propagation vector ${\bm q}_{c ,~ 1 } = \left( \frac{1}{2} ,~ \frac{1}{2} ,~ \frac{1}{6} ~ \text{or} ~ \frac{1}{3} \right) $ and a phase difference of $165 ^{\circ}$ between the Ce(1) and Ce(2) sublattices.
This quantitative result was obtained with the present ``transferred-hyperfine-field'' model.
In addition, the magnetic dipolar model resulted in a magnetic moment at the Ce(2) site that was much larger than 2.54 $\mu_{\rm B} $; in comparison, this model can explain the experimental results as ${\bm m}_{ 2 } (T=0) = 1.72 \mu_{\rm B }$, which is a quantitatively realistic value.
The ratio of the magnitudes of the magnetic moments at the Ce(1) and Ce(2) sites was evaluated as $\frac{ m_{1} }{ m_{2} }\sim \frac{1}{40}$.
This is consistent with the specific heat results.
We conclude that the Ce(2) site plays a dominant role in the magnetism of Ce$_{3}$PtIn$_{11}$.
Concurrently, the magnetic moment at the Ce(1) site, which is very small but finite, may play an essential role in the successive magnetic transitions.
Therefore, the transitions of Ce$_{3}$PtIn$_{11}$ are considered to be a unique property among the Ce$_{n} M_{m}$In$_{3n + 2m}$ family.
Based on the present results, neutron scattering experiments may yield the complete magnetic structure of Ce$_{3}$PtIn$_{11}$.
Further experiments below $ T_{\rm c} $ are also needed, which may confirm the co-occurrence of magnetism and superconductivity.



\vspace{5mm}
\begin{center}

{\bf ACKNOWLEDGMENTS}

\end{center}

\vspace{5mm}

This work was supported by JSPS KAKENHI Grant (No. 18K03505,  No. 19K14644 and No. 21K03439).
The work in Poland was supported by the National Science Centre (Poland) under Research Grant No. 2015/19/B/ST3/03158.



\begin{thebibliography}{22}

 \bibitem{2005coleman} P. Coleman, and A. J. Schofield, Nature, {\bf 433}, 226-229 (2005). 
 \bibitem{2017belitz} D. Belitz and T. R. Kirkpatrick, Phys. Rev. Lett., {\bf 119}, 267202 (2017). 
 \bibitem{2010si} Q. Si, and F. Steglich, Science {\bf 329}, 1161-1166 (2010). 
 \bibitem{2006park} T. Park, F. Ronning, H. Q. Yuan, M. B. Salamon, R. Movshovich, J. L. Sarrao and J. D. Thompson, Nature {\bf 440}, 65-68 (2006). 
 \bibitem{2019das_usa} D. Das, D. Gnida, P. Wi\'{s}niewski, and D. Kaczorowski, Proc. Natl. Acad. Sci. U.S.A. {\bf 116}, 20333 (2019). 
 \bibitem{2017kaluarachchi} U. S. Kaluarachchi, S. L. Bud'ko, P. C. Canfield, and V. Taufour, Nat. Commun., {\bf 8}, 546 (2017). 
 \bibitem{2000hegger} H. Hegger, C. Petrovic, E. G. Moshopoulou, M. F. Hundley, J. L. Sarrao, Z. Fisk, and J. D. Thompson, Phys. Rev. Lett., {\bf 84}, 4986 (2000). 
 \bibitem{2001petrovic} C. Petrovic, P. G. Pagliuso, M. F. Hundley, R. Movshovich, J. L. Sarrato, J. D. Thompson, Z. Fisk, and P. Monthoux, J. Phys.: Condens. Matter, {\bf 13}, L377 (2001). 
 \bibitem{2001kohori} Y. Kohori, Y. Yamato, Y. Iwamoto, T. Kohara, E. D. Bauer, M. B. Maple, and J. L. Sarrao, Phys. Rev. B, {\bf 64}, 134526 (2001). 
 \bibitem{2001kawasaki} S. Kawasaki, T. Mito, G.-q. Zheng, C. Thessieu, Y. Kawasaki, K. Ishida, Y. Kitaoka, T. Muramatsu, T. C. Kobayashi, D. Aoki, S. Araki, Y. Haga, R. Settai, and Y. Onuki, Phys. Rev. B, {\bf 65}, 020504(R) (2001).
 \bibitem{2007fukazawa} H. Fukazawa, T. Okazaki, K. Hirayama, Y. Kohori, G. Chen, S. Ohara, I. Sakamoto, and T. Matusmoto, J. Phys. Soc. Jpn., {\bf 76}, 124703 (2007). 
 \bibitem{2010Kaczorowski} D. Kaczorowski, D. Gnida, A. P. Pikul, and V. H. Tran, Solid State Commun., {\bf 150}, 411 (2010). 
 \bibitem{2010yashima} M. Yashima, S. Taniguchi, H. Miyazaki, H. Mukuda, Y. Kitaoka, H. Shishido, R. Settai, and Y. Onuki, J. Phys.: Conf. Series, {\bf 200}, 012238 (2010). 
 \bibitem{2012fukazawa} H. Fukazawa, R. Nagashima, S. Shimatani, Y. Kohori, and D. Kaczorowski, Phys. Rev. B, {\bf 86}, 094508 (2012). 
 \bibitem{1998mathur} N. D. Mathur, F. M. Grosche, S. R. Julian, I. R. Walker, D. M. Freye, R. K. W. Haselwimmer and G. G. Lonzarich, Nature, {\bf 394}, 39-43 (1998). 
 \bibitem{2001monthoux} P. Monthoux and G. G. Lonzarich, Phys. Rev. B, {\bf 63}, 054529 (2001). 
 \bibitem{1980benoit} A. Benoit, J. X. Boucherle, P. Convert, J. Flouquet, J. Palleau, and J. Schweizer, Solid State Commun., {\bf 34}, 293 (1980). 
 \bibitem{2001bao} W. Bao, P. G. Pagliuso, J. L. Sarrao, J. D. Thompson, Z. Fisk, and J. W. Lynn, Phys. Rev. B, {\bf 64}, 020401(R) (2001). 
 \bibitem{2011sakai} H. Sakai, Y. Tokunaga, S. Kambe, H.-O. Lee, V. A. Sidorov, P. H. Tobash, F. Ronning, E. D. Bauer, and J. D. Thompson, Phys. Rev. B, {\bf 83}, 140408(R) (2011). 
 \bibitem{2017raba} M. Raba, E. Ressouche, N. Qureshi, C. V. Colin, V. Nassif, S. Ota, Y. Hirose, R. Settai, P. Rodiere, and I. Sheikin, Phys. Rev. B, {\bf 95}, 161102(R) (2017).
 \bibitem{2017gauthier} N. Gauthier, D. Wermeille, N. Casati, H. Sakai, R. E. Baumbach, E. D. Bauer, and J.S. White, Phys. Rev. B, {\bf 96}, 064414 (2017).
 \bibitem{2000curro} N. J. Curro, P. C. Hammel, P. G. Pagliuso, J. L. Sarrao, J. D. Thompson, and Z. Fisk, Phys. Rev. B, {\bf 62}, R6100 (2000). 
 \bibitem{2006curro} N. J. Curro, New J. Phys., {\bf 8}, 173 (2006). 
 \bibitem{2014kratochvilova} M. Kratochv\'{i}lov\'{a}, M. Du\v{s}ek, K. Uhl\'{i}\v{r}ov\'{a}, A. Rudajevov\'{a}, J. Prokle\v{s}ka, B. Vondr\'{a}\v{c}kov\'{a}, J. Custers, and V. Sechovsk\'{y}, J. Cryst. Growth, {\bf 397}, 47 (2014). 
 \bibitem{2018das} D. Das, D. Gnida, {\L}. Bochenek, A. Rudenko, M. Daszkiewicz, and D. Kaczorowski, Sci. Rep., {\bf 8}, 16703 (2018).  
 \bibitem{2015prokleska} J. Prokle\v{s}ka, M. Kratochv\'{i}lov\'{a}, K. Uhl\'{i}\v{r}ov\'{a}, V. Sechovsk\'{y}, and J. Custers, Phys. Rev. B, {\bf 92}, 161114(R) (2015). 
 \bibitem{2015kratochvilova} M. Kratochv\'{i}lov\'{a}, J. Prokle\v{s}ka, K. Uhl\'{i}\v{r}ov\'{a}, V. Tk\'{a}\v{c}, M. Du\v{s}ek, V. Sechovsk\'{y}, and J. Custers, Sci. Rep., {\bf 5}, 15904 (2015).
 \bibitem{2019das} D. Das, D. Gnida, and D. Kaczorowski, Phys. Rev. B, {\bf 99}, 054425 (2019).
 \bibitem{2003nicklas} M. Nicklas, V. A. Sidorov, H. A. Borges, P. G. Pagliuso, C. Petrovic, Z. Fisk, J. L. Sarrao, and J. D. Thompson, Phys. Rev. B, {\bf 67}, 020506(R) (2003). 
 \bibitem{2003kawasaki} S. Kawasaki, T. Mito, Y. Kawasaki, G.-q. Zheng, Y. Kitaoka, D. Aoki, Y. Haga, and Y. Onuki, Phys. Rev. Lett., {\bf 91}, 137001 (2003). 
 \bibitem{2011benlagra} A. Benlagra, L. Fritz, and M. Vojta, Phys. Rev. B, {\bf84}, 075126 (2011).
 \bibitem{2020fukazawa} H. Fukazawa, K. Kumeda, N. Shioda, Y. Lee, Y. kohori, K. Sugimoto, D. Das, J. B{\l}awat, and D. Kaczorowski, Phys. Rev. B, {\bf 102}, 165124 (2020). 
 \bibitem{2021das} D. Das, J. B{\l}awat, D. Gnida, and D. Kaczorowski, Physica B: Condensed Matter, {\bf 603}, 412724 (2021). 
 \bibitem{2020kambeprb} S. Kambe, H. Sakai, Y. Tokunaga, R. E. Walstedt, M. Kratochv\'{i}lov\'{a}, K. Uhl\'{i}\v{r}ov\'{a}, and J. Custers, Phys. Rev. B, {\bf 101}, 081103(R) (2020). 
 \bibitem{2000bao} Wei Bao, P. G. Pagliuso, J. L. Sarrao, J. D. Thompson, Z. Fisk, J. W. Lynn, and R. W. Erwin, Phys. Rev. B, {\bf 62}, R14621 (2000). 


\end{thebibliography}
\end{document}